\documentclass[aps,final,notitlepage,oneside,twocolumn,nobibnotes,nofootinbib,
superscriptaddress,noshowpacs,centertags]{revtex4-1}

\usepackage[utf8]{inputenc}
\usepackage[english]{babel}
\usepackage{graphicx}
\usepackage{latexsym}
\usepackage{amssymb}
\usepackage{amsmath}
\usepackage{float}
\usepackage{color}

\begin{document}

\title{
Constraints on the Number of Primordial Black Holes\\
Due to Interaction with Dust
}

\author{A. N. Melikhov}\thanks{e-mail: melikhov94@inbox.ru}
\affiliation{Astro Space Center, P. N. Lebedev Physical Institute, Russian Academy of Sciences, Moscow, 117997 Russia}
\author{E. V. Mikheeva}\thanks{e-mail: helen@asc.rssi.ru}
\affiliation{Astro Space Center, P. N. Lebedev Physical Institute, Russian Academy of Sciences, Moscow, 117997 Russia}

\date{\today}

\begin{abstract}
Photons emitted by primordial black holes (PBHs) due to the Hawking effect are among the factors
of interstellar dust heating. Based on the data on the temperature of dust, constraints on the fraction of $10^{15}\leq M\leq10^{17}$~g.

\vspace{3pt}
\end{abstract}
\maketitle

\section{INTRODUCTION}

Interest in primordial black holes (PBHs) has increased significantly since the recent detection of
gravitational waves from merging black holes by LIGO \cite{1}. The analysis of the data revealed that the intrinsic angular momentum of these black holes is close to zero, which is difficult to explain for astrophysical black holes, but quite logical for PBHs. In addition, the masses of merging black holes detected by LIGO turned out to be much larger than the masses of black holes known from other observational data (for example, from the analysis of X-ray binaries; see, e.g., the
review \cite{2} and the analysis carried out in \cite{3}). Moreover, the question of the carrier of dark matter remains open (see the recent review \cite{4}): sterile neutrinos with a mass of about 3 eV, the existence of which was recently supported by promising results \cite{5,6,7}, cannot
completely solve the dark matter problem. Thus, at the moment such candidates as PBHs are being considered for this role; their cosmological properties make them plausible candidates for cold dark matter. The idea that PBHs can constitute dark matter was first expressed in \cite{8}.

In this paper, we consider PBHs in the range of mass from $10^{15}$ to $10^{17}$~g. PBHs could occur as a result of the gravitational collapse of primordial matter inhomogeneities in the early Universe and may constitute a significant part of dark matter \cite{9,10,11,12,13,14,15}. Born
black holes will have masses on the order of the mass within the horizon at the time of their formation:
$M\sim c^3t/G \simeq 5 \times 10^{-19}(t/10^{-23}s)\,M_{\odot}$, where $c$ is the speed of light, $G$ is the gravitational constant, and $M_{\odot}$ is the mass of the Sun. There are constraints on the PBH fraction in the dark matter obtained from the analysis of the extragalactic and galactic gamma radiation backgrounds (\cite{16,17,18,19} and \cite{20,carrandkuhnel2021,21,22,23,24}, respectively), the microwave background \cite{25,26,27,28,29}, and the
cosmic ray background \cite{30}.

\section{INTERSTELLAR DUST}
Interstellar dust is known to be one of the components of the interstellar medium, along with interstellar gas, interstellar electromagnetic fields, cosmic rays, and dark matter \cite{31}. The mass of interstellar dust is approximately 1\% of the mass of interstellar gas. The formation of dust mainly takes place in the slowly outflowing atmospheres of red dwarf stars, as well as during explosive processes on stars and gas ejection from the galactic nuclei. Dust also forms in
planetary and protostellar nebulae, stellar atmospheres, and interstellar clouds. Under the action of gas flows and radiation pressure, dust particles are carried out into the interstellar medium, where they slow down interacting with the gas and cool down to temperatures of $10-20$~K. As a result, low-volatile molecules from the interstellar gas freeze on the dust grains and form a ``dirty ice'' shell (water molecules mixed with many other molecules). Under the action of electron
sticking and photoionization of dust grains by stellar radiation, dust grains become electrically
charged and therefore are able to interact with electromagnetic fields. An observable manifestation of interstellar dust is its absorption of starlight, as a result of which the light of stars weakens and becomes redder, since the extinction in the optical range is inversely proportional to the wavelength. The emission spectrum of interstellar dust in the infrared and submillimeter frequencies serves as an indicator of physical conditions, and the radiated power can provide information about star populations that cannot be acquired by other means. Interstellar dust takes an active part in the cooling of the interstellar medium and, therefore, contributes to the processes of star formation \cite{31}.

A photon absorbed by a dust grain initiates the thermal motion of the particles of the dust grain. In
this case, the dust grain begins to radiate in a continuous spectrum, which can be approximated by the Planck spectrum of black body radiation. Most of the ultraviolet radiation from stars in the Galaxy is converted into infrared radiation of dust grains.

At present, there is no consensus on the chemical composition and form of interstellar dust. There are several models that explain the properties of interstellar dust. In this paper, we consider the MRN model proposed in \cite{32}. According to this model, interstellar dust consists of a mixture of graphite and silicate particles in approximately equal mass proportions, while the particles have a spherical shape and size $0,005 < a < 0,25 \mu\rm m$, and their size distribution follows
a power law $n(a)\sim a^{-3.5}$. The advantage of this model is that it adequately explains the interstellar extinction curve in the wavelength range $1100-10000\, {\buildrel _{\circ} \over {\mathrm{A}}}$. Figure 1 shows the observational data on interstellar extinction, the theoretical curve following from the MRN model, and the contribution of graphite particles. Graphite particles are responsible for the excess absorption at a wavelength of $2175\, {\buildrel _{\circ} \over {\mathrm{A}}}$.

\begin{figure}
\renewcommand{\baselinestretch}{1}
\centerline{\includegraphics[width=8cm]{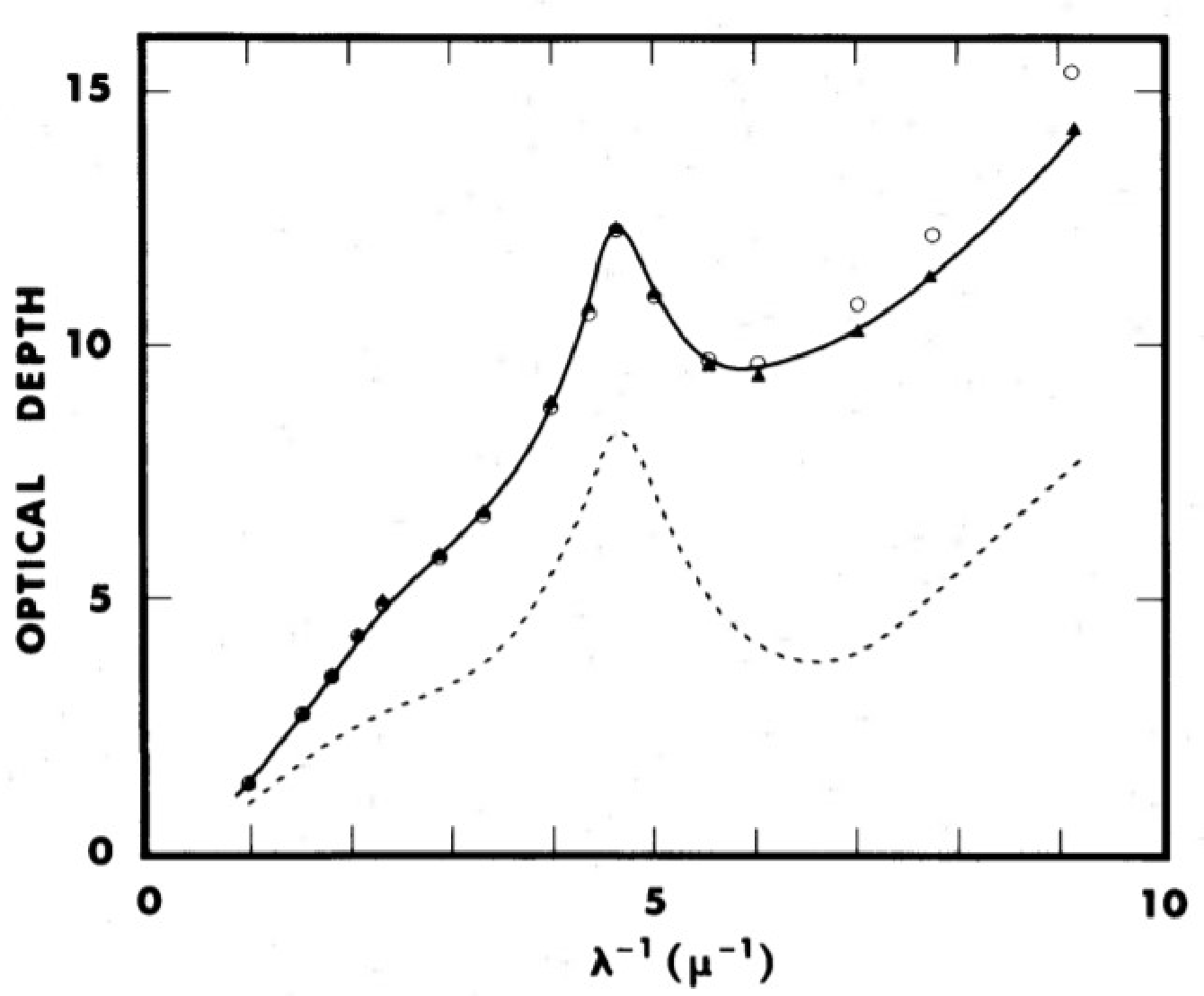}}
\caption{
Optical depth for column density of $10^{22}$ H atoms cm$^{-2}$ as a function of inverse wavelength. The thick line is the curve corresponding to the MRN model, the dots are the experimental data, and the dashed line is the contribution of graphite particles \cite{31}.}
\end{figure}

\section{THERMAL EQUILIBRIUM AND DUST TEMPERATURE}
Dust absorption efficiency plays an important role in the energy equilibrium. Dust absorption efficiency is ratio of the absorption cross section to the geometric cross section of a dust grain:
\begin{equation}
    Q(\lambda)=\frac{C_{abs}}{{\sigma}_d},
\end{equation}
where the geometric cross section of a dust grain $\sigma_d$ equals $\pi a^2$, and $a$ is the dust grain radius.

For estimative calculations, the absorption efficiency is often taken as in \cite{33}
\begin{equation}
    Q(\lambda) =
    \begin{cases}
    1, &\text{$\lambda \leq 2\pi a$}\,,\\
    \frac{2\pi a}{\lambda}, &\text{$\lambda > 2\pi a$}\,.
    \end{cases}
\end{equation}

One of the most important characteristics of the dust is its temperature. The equilibrium temperature of a dust grain is determined from the conditions of the energy balance of heating and cooling of a dust grain. The heating rate of a dust grain is described by the
expression
\begin{equation}
     \dfrac{dE^{abs}}{dt}=4\pi{\sigma}_d\int_0^{\infty}Q(\lambda)J(\lambda)d\lambda,
\end{equation}
where $J(\lambda)$ is the intensity of a wavelength $\lambda$ radiation in which the dust grain is placed \cite{33}. The cooling rate of a dust grain is
\begin{equation}
    \dfrac{dE^{rad}}{dt}=4\pi{\sigma}_d\int_0^{\infty}Q(\lambda)B(T_d,\lambda)d\lambda,
\end{equation}
where $T_d$ is the equilibrium temperature of the dust grain, and $B(T_d, \lambda)$ is the Planck function.

Expressions (3) and (4) are only valid for large grains ($a\geq0.01\mu\rm m$). Small grains are characterized by a low heat capacity, which leads to a sharp increase in the temperature of small dust grains even when small portions of energy are absorbed, and therefore the dust temperature changes abruptly. Between temperature jumps, most small particles cool down to the
temperature of the cosmic microwave background radiation ($2.7$~K). The emission of dust particles occurs mainly when $T_d$ is above the equilibrium \cite{31}. Thus, in our model, we will consider only large dust grains with sizes from 0.01 to 0.25 $\mu\rm m$.

The temperature of the dust varies depending on the region of the Galaxy. Far from the circumstellar envelopes, in the regions of atomic and molecular hydrogen, the dust temperature can drop to $10-20$~K. If the dust is in the HII zones, its temperature lies in the range of $30-200$~K. The highest dust temperatures occur in the circumstellar envelopes. Here, the dust grain temperature reaches $1000-1500$~K \cite{31}. Also, inside dense clouds, where the radiation from stars is greatly weakened, and the dust is heated mainly due to microwave background radiation, the dust temperature can drop to 6~K \cite{33}.

The paper \cite{33} also gives approximate temperatures for the silicate and graphite components of dust depending on the grain size. For the silicate component of dust, the approximate temperature is
\begin{equation}
    T_{sil}=13.6\Big(\frac{1 \mu\rm m}{a}\Big)^{0.06}\,\rm K,
\end{equation}
for the graphite component
\begin{equation}
    T_{gra}=15.8\Big(\frac{1 \mu\rm m}{a}\Big)^{0.06}\,\rm K.
\end{equation}

\section{PBH EMISSION MODEL}
Assuming that PBHs are uniformly distributed in the Universe, and dust grains are uniformly distributed in our Galaxy, we consider PBHs with masses $10^{15}\leq M \leq10^{17}$~g. Dust grains receive photons emitted by PBHs due to the Hawking effect. We assume that photons propagate freely, so the interaction with matter can be neglected. Absorbing energy from PBHs
at all wavelengths, dust grains heat up and radiate as an absolutely black body in the infrared range with an equilibrium temperature $T_d$, which is determined from the thermal equilibrium condition. Calculating the heating rate of a dust grain, we do not consider heating from other sources. By comparing the heating and cooling rates of dust, we calculate a constraint on the fraction of PBHs that constitute dark matter.

According to \cite{34, 35}, the PBH radiation temperature is described by the expression
\begin{equation}
    k_BT = \frac{\hbar c^3}{8\pi GM}, 
\end{equation}
where $k_B$ is the Boltzmann constant, while the energy spectrum of photons from the evaporation of one PBH is given by the formula
\begin{equation}
    \frac{dN_{\gamma}}{dt\,dE} = \frac{\Gamma}{2 \pi\hbar}\Big[\exp\Big(E/k_BT\Big)-1\Big]^{-1},
\end{equation}
where $\Gamma$ is the gray factor, which for high energies takes the form  $\Gamma = \dfrac{27 G^2 M^2 E^2}{\hbar^2 c^6}$ \cite{36}.

\section{RESULTS}
The radiation flux from the PBH as a function of energy and time can be calculated as follows:
\begin{equation}
F(E,t) = \frac{c}{4\pi}u(E,t),
\end{equation}
where $u$ is the energy density, and $t$ is the cosmological time at which the black hole evaporates.

The PBH energy density recorded at the present time point is given by the expression \cite{16, 17, 37}
$$
u_{0} = (1+z)^{-3}u(E,t_{em}) = n_{PBH}(t_0)\int_{0}^{\infty}g(M)dM
$$
$$    
\times\int_{t_{rec}}^{t_0}dt\int_{0}^{\infty}(1+z)^2E_0\frac{dN_{\gamma}}{dt\,dE}\Big(E_0(1+z)\Big)dE =
$$
$$ = \frac{f \rho_{DM}}{M}\int_{0}^{\infty}g(M)dM\int_{z_{rec}}^{z_0}\Big\vert\frac{dt}{dz}\Big\vert dz\int_{0}^{\infty}(1+z)^2
$$
\begin{equation}
\times E_0\frac{dN_{\gamma}}{dt\,dE}\Big(E_0(1+z)\Big)dE
\end{equation}
where $E_0$, $n_{PBH}(t_0)$, and $\rho_{DM}$ are the photon energy, PBH concentration, and dark matter density at the current time point, respectively; $g(M)$ is the PBH mass function; $t_{rec}$ and $z_{rec}$ are the time and redshift of the recombination moment;
$$
\Big\vert\dfrac{dt}{dz}\Big\vert=
$$
\begin{equation}
\dfrac{1}{(1+z)H_0[\Omega_m(1+z)^3+\Omega_{\Lambda}+\Omega_{\gamma}(1+z)^4]^{\frac{1}{2}}},
\end{equation}
where $H_0$ is the Hubble constant; $\Omega_m$, $\Omega_{\Lambda}$, and $\Omega_{\gamma}$ are the cosmological densities of matter, dark energy, and radiation, respectively.

Thus, the radiation flux from the PBH at the present time is
\begin{equation}
F_0=\frac{c}{4\pi}u_0\,.
\end{equation}

Substituting (12) into (3), we derive the heating rate of a dust grain:
\begin{equation}
\frac{dE^{abs}}{dt}=4\pi \sigma_d F_0.
\end{equation}

Since we consider PBHs with masses $10^{15}-10^{17}$~g, the energy of their radiation is $\sim 1-100$~MeV. Even if they are located at $z=z_{rec}$, at the current time point, photons from such PBHs should arrive with energy $\sim 1-100$~keV, which corresponds to X-ray radiation.
Therefore, the wavelengths at which PBHs radiate $\lambda_{PBH}\ll2\pi a$, and to calculate the heating rate of a dust grain, we take $Q(\lambda)=1$.

Upon cooling, a dust grain radiates in the infrared, so for it $\lambda_d > 2\pi a$, and to calculate the cooling rate we take $Q(\lambda) = 2\pi/\lambda$. Substituting $Q(\lambda)$ into (4), we find
the cooling rate of a dust grain:
\begin{equation}
\frac{dE^{rad}}{dt} = 4\pi \sigma_d F_d,
\end{equation}
where $F_d$ is the radiation flux of a dust particle.

We considered two options for the mass function: a monochromatic mass function ($\delta$-function) and the lognormal distribution first proposed in \cite{9}, the probability density of which has the form \cite{krish2006} 
\begin{equation}
g(M) = \frac{1}{\sqrt{2\pi} \sigma M} \exp\Big(\frac{-\log^2(M/\mu)}{2\sigma^2}\Big),
\end{equation}
where $\mu$ and $\sigma$ are the distribution parameters. Normalization
of the lognormal distribution to $f$ is considered in formula (10).

The strongest constraints on the fraction of PBHs that comprise dark matter are obtained if we take the minimum size of a dust grain $a=0.01\mu\rm m$.  For this size, the temperature of the dust of the graphite and silicate components, respectively, is $T_{gra}=17.93$~K and $T_{sil}=20.83$~K. As a result, the following rate of heating of a dust grain by photons of Hawking radiation was obtained:
\begin{equation}
\frac{dE^{abs}}{dt}=1.27\times10^{-11}\times f\times\Big(\frac{10^{15}g}{M}\Big)^3 erg/s.
\end{equation}

The cooling rate for the silicate component of the dust is 
\begin{equation}
\Big(\frac{dE^{rad}}{dt}\Big)_{sil}=2.76\times10^{-14} \, erg/s\,.
\end{equation}

And for the graphite component, the cooling rate is
\begin{equation}
\Big(\frac{dE^{rad}}{dt}\Big)_{gra}=5.84\times10^{-14} \, erg/s\,.
\end{equation}

The constraint on the fraction of PBHs that comprise dark matter can be obtained by comparing the heating and cooling rates of dust from the assumption that the heating rate should be smaller than the cooling rate. We consider dust temperatures indicated in (5) and (6).

Figure 2 shows the results for the monochromatic mass function, i.e., the upper limit on the PBH fraction in the cosmological density of dark matter $f$ depending on mass $M$. These constraints can be compared with the ones obtained from extragalactic and galactic backgrounds \cite{16}, \cite{30}, \cite{23}, \cite{22}. The constraints obtained in our analysis for the monochromatic mass function turned out to be much weaker.
\begin{figure}
\renewcommand{\baselinestretch}{1}
\centerline{\includegraphics[scale=0.7]{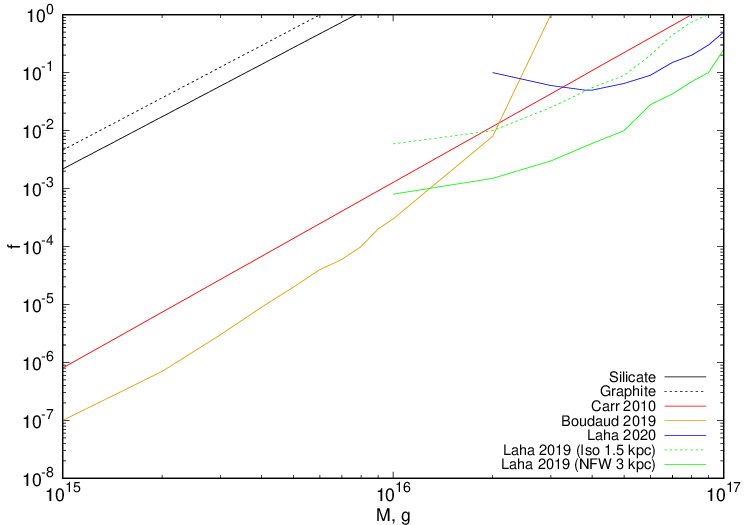}}
\caption{Constraints on the fraction of PBHs that comprise dark matter for the monochromatic mass function. The solid black line shows the constraints if the radiation from the PBH is absorbed by the silicate component of the dust, and the dotted black line shows the same for the graphite component. The constraints obtained from extragalactic and galactic gamma-ray backgrounds
in \cite{16}, \cite{30}, \cite{23}, \cite{22} are given as well (colored lines). 
}
\end{figure}
 
Figure 3 shows the results for the lognormal distribution and the upper limit on the fraction of PBH $f$
depending on the value $\mu$. The figure also shows the constraints previously obtained by other authors from the extragalactic and galactic radiation backgrounds for the lognormal distribution at the parameter $\sigma = 2$ \cite{22}, \cite{30}, \cite{carrandkuhnel2021}. For this $\sigma$ value, the constraints obtained by this method turned out to be weaker than in \cite{carrandkuhnel2021}
(based on the data on the gamma background), but stronger than in \cite{22} and \cite{30} at $\mu \sim 10^{15}-3\times10^{16}$ g.
\begin{figure}
\renewcommand{\baselinestretch}{1}
\centerline{\includegraphics[scale=0.7]{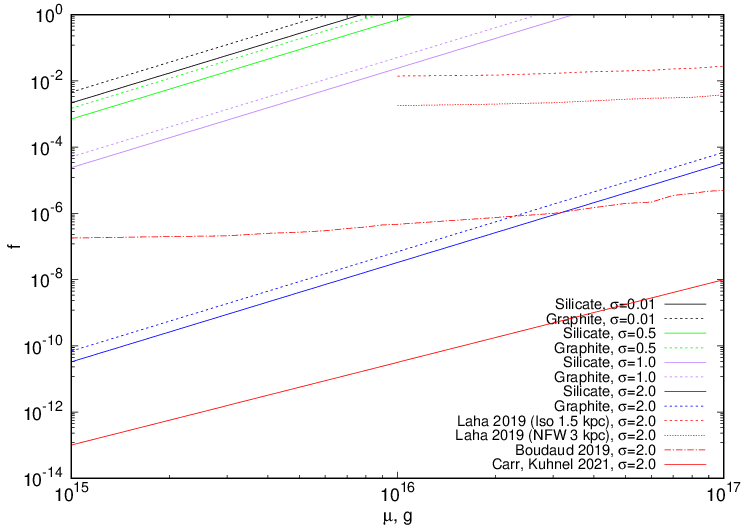}}
\caption{
Constraints on the fraction of PBHs that comprise dark matter for the lognormal distribution. The solid lines show the constraints if the radiation from the PBH is absorbed by the silicate component of the dust, and the dotted lines show the same for the graphite component. For comparison, we present the results obtained for the extragalactic and galactic backgrounds in \cite{22}, \cite{30}, \cite{carrandkuhnel2021} for the lognormal distribution at $\sigma = 2$ (red lines).
}
\end{figure}

\section{CONCLUSIONS}

The importance of studying PBHs lies in the fact that they can be important in explaining various phenomena, from dark matter to the formation of supermassive black holes. The variety of phenomena in which PBHs can participate is due to the wide range of their masses. However, it should be considered that the physical effects of PBHs should not contradict the measured effects.

In this study, the process of heating of dust grains by PBHs uniformly filling the Universe was examined for the first time. The contribution of other radiation sources to dust heating was not considered. The monochromatic mass function and the lognormal distribution were considered in the study. The constraints obtained here for the monochromatic mass function turned out to be weaker than in previous studies, where the constraints were obtained from the PBH
contribution to the gamma background. For the lognormal PBH mass distribution, the constraints turned out to be stricter than in \cite{22} and \cite{30}, but less strict than in \cite{carrandkuhnel2021} for the same value $\sigma = 2$.

\section*{ACKNOWLEDGMENTS}
The authors thank V.N. Lukash and P.B. Ivanov for the review of the paper and their remarks, as well as the referee for comments and suggestions.

\section*{FUNDING}
The study was supported by the Russian Foundation for Basic Research, grant no. 19-02-00199.

\section*{CONFLICT OF INTEREST}
The authors declare that they have no conflicts of interest.

\end{document}